\documentclass[aps,10pt,prl,bibliography,citeautoscript,superscriptaddress,twocolumn]{revtex4-2}
\usepackage[T1]{fontenc}
\usepackage{graphicx}
\usepackage{dcolumn}
\usepackage{amsmath,amssymb,amsfonts,bm,dsfont,units}
\usepackage{hyperref}
\hypersetup{colorlinks=true,citecolor=blue,linkcolor=blue,urlcolor=blue,pdfstartview=FitH}
\usepackage{braket}
\usepackage{float,latexsym}
\usepackage{multirow}
\usepackage[normalem]{ulem} % for \sout
\usepackage[caption=false]{subfig} % for subfloat
\newcommand{\sfigure}[2]{Figure~\hyperref[#1]{\ref{#1}(#2)}}
\newcommand{\sfigref}[2]{Fig.~\hyperref[#1]{\ref{#1}(#2)}}

\usepackage{color}
\definecolor{dkgreen}{rgb}{0,0.5,0}
\definecolor{midnightblue}{rgb}{0.39,0.58,0.93}

\definecolor{kspink}{RGB}{200,0,200}

\newcommand{\comment}[1]{}{}
\begin{document}

\title{Dynamical anyon generation in Kitaev honeycomb non-Abelian spin liquids}
	\author{Yue Liu}
\affiliation{Department of Physics and Institute for Quantum Information and Matter, California Institute of Technology, Pasadena, CA 91125, USA}
	\author{Kevin Slagle}
\affiliation{Department of Physics and Institute for Quantum Information and Matter, California Institute of Technology, Pasadena, CA 91125, USA}
\affiliation{Walter Burke Institute for Theoretical Physics, California Institute of Technology, Pasadena, CA 91125, USA}
    \author{Kenneth S. Burch}
\affiliation{Department of Physics, Boston College, Chestnut Hill, Massachusetts 02467, USA}
    \author{Jason Alicea}
\affiliation{Department of Physics and Institute for Quantum Information and Matter, California Institute of Technology, Pasadena, CA 91125, USA}
\affiliation{Walter Burke Institute for Theoretical Physics, California Institute of Technology, Pasadena, CA 91125, USA}

\date{\today}

	% Activate to display a given date or no date
	
	\begin{abstract}
		Relativistic Mott insulators known as `Kitaev materials' potentially realize spin liquids hosting non-Abelian anyons. 
		Motivated by fault-tolerant quantum-computing applications in this setting, we introduce a dynamical anyon-generation protocol that exploits universal edge physics.  The setup features holes in the spin liquid, which define energetically cheap locations for non-Abelian anyons, connected by a narrow bridge that can be tuned between spin liquid and topologically trivial phases.  We show that modulating the bridge from trivial to spin liquid over intermediate time scales---quantified by analytics and extensive simulations---deposits non-Abelian anyons into the holes with $\mathcal{O}(1)$ probability.  The required bridge manipulations can be implemented by integrating the Kitaev material into magnetic tunnel junction arrays that engender locally tunable exchange fields. 
		Combined with existing readout strategies, our protocol reveals a path to topological qubit experiments in Kitaev materials at zero applied magnetic field.
	\end{abstract}
	\maketitle

{\bf \emph{Introduction.}}~The Kitaev honeycomb model captures an exactly solvable, gapless spin liquid that serves as a parent phase for nearby gapped topological orders \cite{KITAEV20062}.  Most strikingly, a descendant gapped spin liquid supporting non-Abelian anyons---the workhorse of intrinsically fault-tolerant quantum computation \cite{Kitaev2003,Nayak08}---emerges upon breaking time-reversal symmetry.  Prospects for laboratory realization rose following the ingenious proposal \cite{Jackeli09} that spin-orbit-coupled Mott insulators now known as Kitaev materials \cite{Savary17,Winter17,Trebst17,Hermanns18,Janssen19,Takagi19,Motome20} exhibit dominant spin interactions of the type present in the Kitaev model.  Among such materials, $\alpha$-RuCl$_3$ has generated particular attention given extensive evidence for fractional excitations \cite{Wang2020KitRam,banerjee2017neutron,WangLoidl2017} and recent thermal transport measurements that possibly indicate the onset of a magnetic-field-driven non-Abelian spin liquid \cite{Kasahara18,Kasahara18nat,Takagi19,Yokoi21,Bruin21}.  While the experimental situation remains to be fully settled \cite{Bachus20,Yamashita20,Bachus21,Chern21,Czajka21}, these results strongly motivate pursuing Kitaev materials as a  venue for eventual quantum information applications.

Exploiting Kitaev materials for fault-tolerant quantum computation requires the development of practical techniques, tailored to an electrically inert platform, for single-anyon detection as well as controlled generation and manipulation of anyons.  Numerous anyon detection methods have recently been proposed in this context, relying on either variations of anyon interferometry \cite{PhysRevX.10.031014,PhysRevLett.126.177204,klocke2021thermal,PhysRevLett.127.167204}
envisioned originally for quantum Hall platforms \cite{PhysRevLett.94.166802,Stern06,Bonderson06} or local probes such as scanning tunneling microscopy \cite{Feldmeier20,Pereira20,Konig20,Udagawa21,Kishony21}.  The prevailing strategy for anyon generation pursued so far seeks perturbations that locally remove the excitation energy for anyons---thus forcing them into the system's ground state at prescribed locations. Near the exactly solvable point of the Kitaev honeycomb model, for instance, atomic-scale perturbations (including impurity spins and vacancies) have been shown to energetically favor the formation of gauge fluxes that constitute Ising  non-Abelian anyons \cite{Willans10,Willans11,PhysRevLett.117.037202,PhysRevX.11.011034,Jang21}. By its nature, this approach requires  sufficiently accurate modeling of an experimental candidate's microscopic Hamiltonian to enable reliable predictions for energetics. Experimental implementation would additionally require exquisite material control and manipulation. 

\begin{figure}
    %\captionsetup{singlelinecheck = false, justification=raggedright}
	\includegraphics[width=\linewidth]{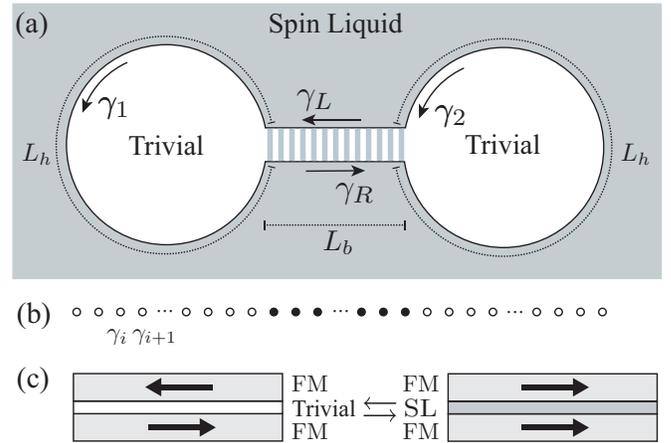}
	
	\caption{\label{fig:geometry} (a) Dumbbell setup used for anyon generation.  A non-Abelian spin liquid hosts two trivial holes connected by a bridge (hatched region).  Evolving the bridge interior from a trivial phase to a spin liquid over a time scale $\tau$ [satisfying Eq.~\eqref{eq:T}] deposits Ising anyons (with appreciable probability) into the adjacent holes, without generating unwanted excitations. 
	(b) Sketch of lattice model used to simulate the spin-liquid protocol.  Solid and open circles respectively represent Majorana fermions that emulate the bridge and holes.
	(c) Sketch of magnetic tunnel junctions that enable the required dynamical manipulations.  Ferromagnetic metals encasing a local patch of the Kitaev material contribute exchange fields that add in the parallel configuration (producing the spin liquid) or cancel in the antiparallel configuration (producing magnetic order).
	}
\end{figure}

We introduce a complementary scheme that  generates Ising anyons as long-lived excitations above the ground state via a dynamical protocol that relies on universal edge physics rather than detailed microscopics.  \sfigure{fig:geometry}{a} illustrates the required setup, consisting of a non-Abelian spin liquid with two holes connected by a narrow bridge.  The holes are always in a topologically trivial phase (e.g., vacuum or magnetically ordered) and thus host a chiral Majorana edge mode at their boundary.  With large enough hole perimeter, Ising anyons become the cheapest edge excitation and can be created by dynamically modulating the bridge.  We specifically assume that the bridge can be evolved over a time scale $\tau$ from a trivial phase (yielding additional chiral Majorana edge states connecting the two holes) to a spin liquid phase (yielding disconnected holes).  Using analytical arguments and extensive numerical simulations, we show that there exists a broad window of $\tau$ such that this evolution deposits an Ising anyon in each hole with $\mathcal{O}(1)$ probability---\emph{without} generating spurious excitations in the bridge.  Once created, the Ising anyons are prevented from re-combining by the bulk energy gap in the surrounding regions.  

After developing our protocol in generality, we propose an implementation scheme that replaces the applied magnetic field traditionally used to form a non-Abelian phase with locally tunable ferromagnets exchange-coupled to the Kitaev material [\sfigref{fig:geometry}{c}]. Local regions of the Kitaev material could be toggled in and out of the spin liquid by controlling the relative orientation of the adjacent ferromagnetic moments---thereby enabling real-time manipulation of the bridge and holes. Together with existing anyon-detection strategies, our anyon generation protocol reveals a possible pathway to fusion and braiding experiments in non-Abelian spin liquids. 

{\bf \emph{Setup and model.}}~Non-Abelian Kitaev spin liquids host a gapless chiral Majorana edge mode described by a chiral Ising conformal field theory (CFT) with central charge $c = 1/2$ \cite{AppliedCFT}.  The bulk supports  three types of gapped quasiparticles: bosons (labeled $I$), emergent fermions $(\psi)$, and non-Abelian Ising anyons ($\sigma$) that carry Majorana zero modes.  Although bulk quasiparticle excitation energies depend sensitively on microscopic details, their edge counterparts display universal low-energy properties dictated by the CFT.  In particular, an Ising anyon dragged to the edge changes the boundary conditions for the chiral Majorana fermions from antiperiodic to periodic, thereby incurring an energy cost $E_\sigma = \frac{1}{16}\frac{2\pi v}{L}$ \cite{Bloete1986,Affleck1986,Feiguin2007,MilstedCFT} with $v$ the edge velocity and $L$ the edge perimeter  (we set $\hbar = 1$ throughout).  Edge Majorana fermions in turn carry energy $E_\psi = \frac{2\pi v}{L}p$, where $p$ is a half-integer for anti-periodic boundary conditions and integer for periodic boundary conditions; in the latter case the $p = 0$ level is the Majorana zero mode bound to an Ising anyon.  Bosonic excitations arise from adding an even number of edge fermions.  

Consider now the ``dumbbell'' geometry of \sfigref{fig:geometry}{a} containing holes of circumference $L_h$ connected by a bridge of length $L_b$.
Since the low-energy physics occurs only on the boundary (if the bridge is sufficiently narrow), we can model the relevant dynamics via an effective Hamiltonian
\begin{equation}
    \mathcal{H} = \mathcal{H}_{h,1} + \mathcal{H}_{b} + \mathcal{H}_{h,2}
    \label{Hfull}
\end{equation}
for the dumbbell edge modes.
Here 
\begin{equation}
    \mathcal{H}_{h,n} = \int_{0}^{L_h} dx \left( -i v\gamma_{n} \partial_x \gamma_{n}\right), \quad n=1,2
\end{equation}
captures the kinetic energy for Majorana fermions $\gamma_1$ and $\gamma_2$ at the left and right holes, respectively.  
The term $\mathcal{H}_b$ governs the left- and right-moving Majorana fermions $\gamma_{L,R}$ residing across the bridge.  Crucially, these modes may be either gapless or fully gapped depending on whether the bridge realizes a trivial or spin liquid phase.  Both regimes are accessible from the interacting bridge Hamiltonian
\begin{align}
    \mathcal{H}_{b} = \int_{0}^{L_b} dx [&-i v\gamma_R \partial_x \gamma_R + i v\gamma_L \partial_x \gamma_L 
    \nonumber \\
    &-\kappa  (\gamma_R \partial_x \gamma_R)(\gamma_L \partial_x \gamma_L)].
    \label{Hb}
\end{align}
Field operators appearing in the Hamiltonian must be continuous at the bridge/hole boundaries, e.g., $\gamma_1(0) = \gamma_L(0)$, $\gamma_1(L_h) = \gamma_R(0)$, etc.

In the limit $\kappa = 0$, $\mathcal{H}_b$ simply encodes the kinetic energy for decoupled right- and left-movers, as appropriate when the bridge is trivial. Here the bridge links the two holes, and the entire dumbbell can be treated as a single chiral Majorana mode traversing a loop of length $L = 2L_h + 2L_b$.   The $\kappa$ interaction on the second line represents the leading local process that couples right- and left-movers near this limit (single-fermion backscattering processes are forbidden since only bosons can tunnel across the trivial bridge) \cite{PhysRevX.10.031014,Lichtman21}.  At weak coupling $\kappa$ is irrelevant and yields only perturbative corrections at low energies.  

As the bridge morphs from trivial to spin liquid,  $\kappa$ increases and drives the bridge boundary from a $c = 1/2$ Ising CFT to a $c = 7/10$ tricritical Ising (TCI) CFT, and then catalyzes spontaneous mass generation \cite{Rahmani16,PhysRevLett.115.166401,PhysRevLett.120.206403,Rahmani2019}
that gaps out the right- and left-movers.  In the gapped phase, the bridge Hamiltonian admits a simple mean-field decomposition:
$\mathcal{H}_{b} \rightarrow \int_x (-i v\gamma_R \partial_x \gamma_R + i v\gamma_L \partial_x \gamma_L + i m \gamma_R \gamma_L)$,
with $m$ the spontaneously generated mass that signals gap formation.  Here the two holes in the dumbbell decouple at low energies---as appropriate when the bridge is spin liquid---and to a good approximation realize independent chiral Majorana modes each propagating over a length $L_h$.  

The boundary conditions for the decoupled Majorana modes nevertheless depend on the sign of the spontaneously generated mass.  To appreciate this point, note first that the local energy in the bridge region can not depend on the sign of $m$ since the mass is generated spontaneously.  Kinks at which the mass changes sign do, however, cost energy; such excitations bind Majorana zero modes \cite{Teo14} and thus correspond to gapped Ising anyons localized in the bridge.  Pulling a kink-antikink pair out of the vacuum and then dragging them to opposite ends of the bridge thereby globally changes the sign of the mass---and deposits a single Ising anyon to each hole that toggles the Majorana fermion boundary conditions as described above.  

We label eigenstates of the decoupled holes by $a_1 \times a_2$, where $a_j$ is the anyon charge in hole $j$.  
We assume that the dumbbell has trivial total topological charge, so that the ground state corresponds to $I\times I$ while the first excited state corresponds to $\sigma \times \sigma$ with excess energy $E_{\sigma\times \sigma} = 2 \times \frac{1}{16} \frac{2\pi v}{L_h}$.  Further excited states with trivial topological charge arise from adding an even number of fermions to the boundary.  
Importantly, the excitation energy for the $\sigma \times \sigma$ state dwarfs that of the next accessible excited state by nearly an order of magnitude---facilitating targeted dynamical creation of Ising anyons as described below.

{\bf \emph{Dynamical anyon-generation protocol.}}~Our protocol begins with the bridge in a trivial phase and a single chiral Majorana mode encircling the dumbbell initialized into its $I\times I$ ground state.
Next, over a time scale $\tau$ we evolve the bridge into a spin liquid phase---thus increasing $\kappa$ in Eq.~\eqref{Hb} until $\gamma_{L,R}$ are fully gapped and the holes decouple.  If $\tau$ is too short, then the evolution will generate unwanted excitations in the bridge region.  If $\tau$ is too long, then the system simply follows adiabatic evolution into the $I\times I$ ground state.  We seek intermediate $\tau$ such that the system lands in the local ground state of the bridge but exhibits a superposition of $I\times I$ \emph{and} $\sigma \times \sigma$ states.  Measurement of the anyon charge at one of the holes then collapses the wavefunction into a well-defined anyon sector; the protocol resets and repeats until measurement returns the desired $\sigma\times \sigma$ state.  

We can obtain an order-of-magnitude estimation of the desired window for $\tau$ using Landau-Zener-type reasoning \cite{Landau32_1,Landau32_2,Zener32}.  Since our protocol modifies only the bridge Hamiltonian, it is useful to temporarily neglect the holes (e.g., by taking $L_h=0$).  In this case the bridge encounters a minimal gap of order $v/L_b$
en route to attaining its final, maximal gap (comparable to the bulk gap $\Delta_{\rm bulk}$) at time $\tau$.  The probability for accessing bridge excited states---either quasiparticles that increase the bridge's final bulk energy density, or virtual kink-antikink pairs that mediate formation of $\sigma \times \sigma$---occur predominantly over a `transition time' \cite{PhysRevLett.62.2543} $\tau_{\star} \sim [(v/L_b)/\Delta_{\rm bulk}]\tau$ around the minimal gap.
The probability of accessing a level with energy $\sim\omega$ during this interval becomes appreciable when $\omega \tau_{\star} \lesssim 1$.  Final states exhibiting (unwanted) bridge excitations have $\omega \gtrsim v/L_b$; avoiding such states thus requires $\tau \gg (L_b/v)^2 \Delta_{\rm bulk}$.  To assess the probability for the targeted $\sigma \times \sigma$ state, we now restore the holes, whose key role is to modify the $\sigma$ excitation energy from the bulk value to $\omega \sim v/L_h$.  Correspondingly, we expect to access $\sigma\times \sigma$ with appreciable probability provided $\tau \lesssim (L_b/v)(L_h/v) \Delta_{\rm bulk}$.  In summary, the time scale $\tau$ should satisfy
\begin{equation}\label{eq:T}
      \left(\frac{L_b}{v}\right)^2 \Delta_{\rm bulk} \ll \tau \lesssim \left(\frac{L_b}{v}\right)\left(\frac{L_h}{v}\right)\Delta_{\rm bulk},
\end{equation}
which always admits a permissible $\tau$ range if $L_b\ll L_h$.  We will bolster Eq.~\eqref{eq:T} by numerically simulating the dynamical evolution in an effective lattice model.

{\bf \emph{Effective lattice model.}}~Directly simulating the protocol dynamics using the interacting continuum model in Eq.~\eqref{Hfull} poses a nontrivial technical challenge.  We instead study a lattice model that yields the same low-energy behavior yet is amenable to large-scale numerics.  Imagine vertically flattening the holes in \sfigref{fig:geometry}{a} so that the dumbbell turns into a line hosting \emph{non-chiral} Majorana fermions over an effective total length $L_{\rm eff} = L_h/2 + L_b + L_h/2$. The same  non-chiral degrees of freedom can emerge from an interacting variant of the Kitaev chain \cite{PhysRevX.10.031014} describing lattice Majorana fermions $\gamma_{1,\ldots,L_{\rm eff}}$; see \sfigref{fig:geometry}{b}, where sites indicated by open and solid circles respectively mimic the holes and bridge.  

We specifically consider
\begin{equation}
    H = H_0 + H_{\rm int} + \delta H.
    \label{Heff}
\end{equation}
The first term, 
\begin{equation}
    H_0 = iJ\sum_{j = 1}^{L_{\rm eff}-1} \gamma_j \gamma_{j+1},
\end{equation}
describes the usual Kitaev chain Hamiltonian tuned to the transition between the trivial phase and topological phase hosting boundary Majorana zero modes.  The low-energy degrees of freedom are massless Majorana fermions $\gamma_{R/L}$, obtained by expanding $\gamma_j \sim \gamma_L + (-1)^j \gamma_R$, with velocity $v = 4J$.    Vanishing of the mass is guaranteed by the single-site Majorana translation symmetry $\gamma_j \rightarrow \gamma_{j+1}$ built into $H_0$, modulo boundary effects.  The second term introduces four-fermion interactions \cite{PhysRevLett.120.206403} among the central $L_b$ sites in the chain [solid circles in \sfigref{fig:geometry}{b}]:
\begin{equation}
  H_{\rm int} = \lambda \sum_{j = \frac{L_h}{2} + 3}^{\frac{L_h}{2} + L_b - 2}\gamma_{j-2}\gamma_{j-1}\gamma_{j+1}\gamma_{j+2}.
  \label{Hint}
\end{equation}
Single-site Majorana translation symmetry continues to preclude explicit mass generation except at the left and right endpoints of the interacting $L_b$ sites, where strong translation symmetry breaking generically produces a finite local mass term.  In the bulk of the $L_b$ region, $H_{\rm int}$ generates the four-fermion interactions from Eq.~\eqref{Hb} with $\kappa \propto \lambda$ in the low-energy limit \cite{PhysRevLett.120.206403}.  Finally, 
\begin{equation}
    \delta H = -i \delta J \left(\gamma_{\frac{L_h}{2} + 1} \gamma_{\frac{L_h}{2} + 2} + \gamma_{\frac{L_h}{2} + L_b - 1} \gamma_{\frac{L_h}{2} + L_b} \right)
    \label{Hboundary}
\end{equation}
acts at the endpoints of the $L_b$ region and counteracts the explicitly generated mass---which is unphysical in the spin liquid problem of interest.

At $\lambda = \delta J = 0$, the entire chain is gapless and emulates the trivial-bridge spin liquid setup.  Upon turning on $\lambda$, this gapless phase survives until $\lambda = \lambda_{\rm TCI} \approx 0.428J$, at which the interacting $L_b$ sites realize a TCI CFT \cite{PhysRevLett.120.206403}.  For $\lambda > \lambda_{\rm TCI}$, the $L_b$ region becomes gapped due to spontaneous mass generation---emulating the regime where the bridge is a spin liquid.  The gapped $L_b$ sites admit a particularly simple description at $\lambda = 0.5J$, as the interacting Hamiltonian becomes frustration-free (again modulo boundary effects) in that limit.  In particular, the $L_b$ sites realize zero-correlation-length ground states of either the trivial \emph{or} topological phase of the non-interacting Kitaev chain \cite{PhysRevLett.120.206403}---both of which yield the same energy density away from the endpoints and the same local gap $\Delta_{\rm bulk} \approx 0.55J$ \footnote{$\Delta_{\rm bulk}$ is extracted from DMRG simulations for a periodic chain with $L_h =0$, and depends only weakly on $L_b$ for the values used in our protocol simulations.}.  We associate the trivial sector with $I\times I$ and the topological sector, given its accompanying end Majorana zero modes, with $\sigma \times \sigma$.  Since these sectors yield different boundary conditions for the decoupled `hole' sites on either end, their overall energies differ.  We fix $\delta J = \alpha \lambda$ in Eq.~\eqref{Hboundary} with the coefficient $\alpha = 0.284$ chosen such that the $\sigma \times \sigma$ excitation energy scales like $1/L_h$ at $\lambda = 0.5J$. 

\begin{figure}
	\centering
    \includegraphics[width=\linewidth]{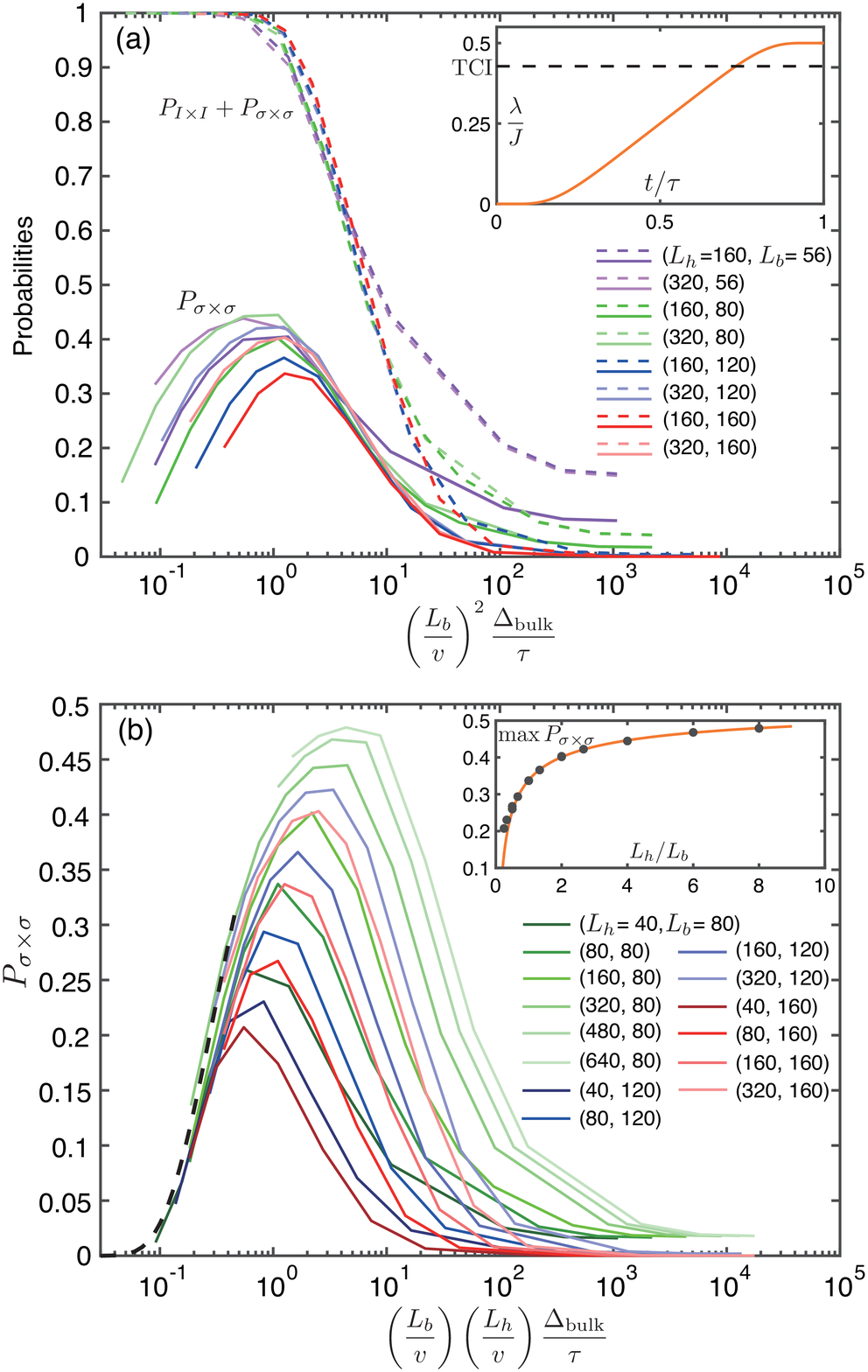}
\caption{\label{fig:TEBD} 
Protocol simulation results. (a) The initial decay of probabilities $P_{I\times I} + P_{\sigma\times\sigma}$ and $P_{\sigma \times \sigma}$ collapses well for all system sizes shown when plotted versus $(L_b/v)^2 \Delta_\text{bulk}/\tau$, supporting the left side of Eq.~\eqref{eq:T}.  Inset: $\lambda$ time dependence used in simulations.
(b) The rise in $P_{\sigma \times \sigma}$ collapses well for all system sizes when plotted versus $(L_b/v) (L_h/v) \Delta_\text{bulk}/\tau$, supporting the right side of Eq.~\eqref{eq:T}. The dashed line fits the rise to $P_{\sigma\times\sigma} = 0.6 \exp{(-0.3 \tau v^2/\Delta_{\rm bulk}L_b L_h)}$.  Inset: maximum of $P_{\sigma\times\sigma}$ for each system size versus $L_h/L_b$. The orange curve [${\rm max}~P_{\sigma\times\sigma} = -0.23\left(L_h/L_b\right)^{-0.45} + 0.57$] fits the large-$L_h/L_b$ data.
}
\end{figure}

{\bf \emph{Protocol simulation.}}~To explore the protocol dynamics in our effective lattice model, we endow $\lambda$ with smooth time dependence, taking $\lambda(t) = f(t/\tau)\lambda(\tau)$ with $f(x)\equiv\{\tanh[\tan((2x-1)\pi/2)]+1\}/2$ and $\lambda(\tau) = 0.5J$; see inset of \sfigref{fig:TEBD}{a}.  
We Jordan-Wigner-transform  Eq.~\eqref{Heff} into a deformed Ising spin chain (see Appendix~\ref{app:JW}) that we simulate using ITensor \cite{ITensor}. At $t=0$ (trivial-bridge configuration), we initialize the system into the ground state, obtained by density-matrix renormalization group (DMRG) \cite{WhiteDMRG,MPSReview} calculations; for details of the DMRG energy level structure see Appendix~\ref{app:levels}. Then we use time-evolution block-decimation (TEBD) \cite{Vidal2003,Vidal2004,MPSReview} to time evolve until $t = \tau$ (spin-liquid-bridge configuration).
The lowest two $t = \tau$ Hamiltonian eigenstates, corresponding to $I\times I$ and $\sigma \times \sigma$, are also obtained by DMRG and used to calculate the probabilities $P_{I \times I}$ and $P_{\sigma \times \sigma}$ for those eigenstates to occur in the final time-evolved wavefunction.
Throughout the simulation, the truncation error is set to be $10^{-9}$ and the time step is set to 0.005 (in $J=1$ units). We checked that the error due to truncation and the finite time step is small by doubling the truncation error and the time step in a system with $L_h = 320, L_b = 56$, and $\tau = 1200$, which only affected the amplitudes for the $I\times I$ and $\sigma \times \sigma$ states by $< 2\%$.

Figure~\ref{fig:TEBD} illustrates the dependence of the probabilities $P_{I \times I}$ and $P_{\sigma \times \sigma}$ on system size and protocol time $\tau$.  In (a) the data are plotted versus $a\equiv(L_b/v)^2 (\Delta_{\rm bulk}/\tau)$, taking  $\Delta_{\rm bulk} = 0.55J$ here and below.  We observe that $P_{I \times I} + P_{\sigma \times \sigma}$ is near unity for $a\lesssim 1$, indicating that the time-evolved wavefunction resides almost entirely in the $I\times I$ and $\sigma \times \sigma$ states, but decays for $a \gtrsim 1$ due to leakage of probability weight into higher excited states.   As $a$ increases, $P_{\sigma \times \sigma}$ initially rises as the protocol escapes the adiabatic regime, but eventually also decays as probability weight shifts towards higher excited states.  The approximate collapse of both $P_{I \times I} + P_{\sigma \times \sigma}$ and $P_{\sigma \times \sigma}$ during the initial descent at $a\gtrsim 1$---for all system sizes---agrees with the left side of Eq.~\eqref{eq:T}.   
During the initial rise in $P_{\sigma \times \sigma}$, by contrast, the curves in (a) certainly do not collapse, i.e., the escape from the adiabatic regime is not set by the parameter $a$. \sfigref{fig:TEBD}{b} plots $P_{\sigma \times \sigma}$ versus $(L_bL_h/v^2) (\Delta_{\rm bulk}/\tau)$.  Excellent data collapse is now observed during the rise---consistent with the right side of Eq.~\eqref{eq:T}.  
The maximum of $P_{\sigma \times \sigma}$ for each system size scales with $L_h/L_b$, as illustrated in the inset of Fig.~\ref{fig:TEBD}; for our protocol the $\sigma \times \sigma$ probability asymptotes at large $L_h/L_b$ to near 1/2.

{\bf \emph{Implementation blueprint.}}~Non-Abelian spin liquid signatures were reported in $\alpha$-RuCl$_3$ over a field interval beginning at $\sim 7$T; at zero field, by contrast, magnetic order appears \cite{Kasahara18,Kasahara18nat,Takagi19,Yokoi21,Bruin21}.  Guided by these observations, we expect that \emph{locally} changing the Zeeman field from `large' to `small' can selectively convert different parts of a Kitaev material between spin liquid and topologically trivial phases, as required for our protocol.   We propose implementing such local variations by forming magnetic tunneling junctions
wherein a monolayer Kitaev material is sandwiched by ferromagnetic metals [\sfigref{fig:geometry}{c}].  Each adjacent ferromagnet induces an exchange field in the Kitaev material.  We assume that in the parallel configuration, the exchange fields from the two layers add to give a net Zeeman field required to form the spin liquid; in the antiparallel configuration, cancellation of the exchange fields instead produces magnetic order in the Kitaev material.  Conversion between parallel and antiparallel configurations can be generated using spin-transfer torque \cite{Ralph08} with nanosecond switch times \cite{Devolder08,Cui10,Devolder16,Devolder16_2} (or even sub-nanosecond with related techniques \cite{Grimaldi20}). 
This approach eschews the need for large external magnetic fields and potentially enables real-time manipulation of holes, bridges, and edge states at the nanoscale.  

Let us estimate the rough length, time, and temperature scales needed for bridge manipulation in our protocol.  To avoid crossing a two-dimensional phase transition on passing to the trivial phase, the bridge thickness should not greatly exceed the bulk spin-liquid correlation length $\xi_{\rm bulk} \sim v_{\rm bulk}/\Delta_{\rm bulk}$, where $v_{\rm bulk}$ is the Dirac velocity for bulk emergent fermions.  For Kitaev couplings $K \sim 8$meV \cite{Sandilands16,Winter17_2,Sears20,Suzuki21} and lattice constant $a \sim 0.6$nm \cite{Johnson15}, the velocity is $v_{\rm bulk} \approx \sqrt{3}K a/4 \sim 3\times 10^3$m/s \cite{Kitaev2003}; taking $\Delta_{\rm bulk} \sim 5$K \cite{Kasahara18} then yields $\xi_{\rm bulk} \sim 5$nm.
The bridge length $L_b$, however, must exceed $\xi_{\rm bulk}$ so that Ising anyons trapped in the holes decouple in the spin-liquid bridge configuration and thus cannot annihilate.  With $L_b \sim 20$nm and a hole perimeter $L_h \sim 200$nm, the above criterion holds while also yielding a $\tau$ window satisfying Eq.~\eqref{eq:T}.  Our protocol then generates Ising anyons with appreciable probability for $\tau \sim \left(\frac{L_b}{v}\right)\left(\frac{L_h}{v}\right)\Delta_{\rm bulk} \sim 3$ns, where we assumed $v \sim 10^3$m/s.  
Finally, since our protocol initializes the system into the ground state of the trivial-bridge configuration, one might anticipate that temperature $T$ must fall below the trivial-bridge excitation energy $\sim v/(2L_h + 2L_b)$.  Ground-state initialization is, however, unnecessary \cite{Akhmerov10} provided the dumbbell remains in the trivial total topological charge sector and the bridge does not trap spurious excitations.  Both conditions are expected to hold for $T$ smaller than the minimal local bridge excitation energy $\sim v/L_b \sim 0.4$K encountered during the protocol (a much milder requirement).

{\bf \emph{Outlook}}.~We proposed a universal, dynamical anyon-generation protocol that employs tunable proximate ferromagnets to probabilistically nucleate Ising anyons in Kitaev materials.    In the future, it would be interesting to optimize the protocol to further boost the achievable $\sigma \times \sigma$ nucleation probability, and also to search for ferromagnets that impart Zeeman couplings needed for non-Abelian spin liquid formation in Kitaev materials such as $\alpha$-RuCl$_3$.  Developing concrete qubit designs and manipulation protocols that combine dynamical anyon generation with interferometric readout appears timely as well.  More broadly, the techniques that we introduced may be profitably adapted to other platforms, including proximitized magnetically doped topological insulators and fractional quantum Hall states.

%%%%%%%%%%%%%%%%%%%%%%%%%%%%%%%%%%%%%%%%%%%%%%%%
\begin{acknowledgments}
It is a pleasure to thank Dave Aasen, Arnab Banerjee, Gabor Halasz, Erik Henriksen, Kai Klocke, Joel Moore, and Ady Stern for stimulating conversations.
This work was also supported by
the U.S. Department of Energy, Office of Science, National Quantum Information Science Research Centers, Quantum Science Center; the Office of Naval Research under Award number N00014-20-1-2308 (KSB);
	the Army Research Office under Grant Award W911NF-17-1-0323; 
	the Caltech Institute for Quantum Information and Matter, an NSF Physics Frontiers Center with support of the Gordon and Betty Moore Foundation through Grant GBMF1250; and
	the Walter Burke Institute for Theoretical Physics at Caltech.
\end{acknowledgments}

\bibliographystyle{apsrev4-2}
\bibliography{anyonGeneration}

\clearpage
% \newpage
\appendix
\section{Jordan-Wagner transformation of effective lattice model}
\label{app:JW}

We map the lattice model in Eq.~\eqref{Heff} to an Ising spin chain using the usual Jordan-Wagner transformation,
\begin{equation}
    \gamma_{2 i-1}=Z_{i} \prod_{j=1}^{i-1} X_{j}, \quad \gamma_{2 i}=-i X_{i} \gamma_{2 i-1},
\end{equation}
where $X$'s and $Z$'s are Pauli matrices
  and $\{\gamma_i, \gamma_j\} = 2 \delta_{i,j}$.
The resulting spin Hamiltonian is a critical transverse-field Ising model supplemented by three-spin interactions in the bridge and boundary transverse fields at the bridge endpoints:
\begin{align}
\label{eq:JW}
    &~H_0 = -J \sum_{j=1}^{\frac{L_{\rm eff}}{2} - 1} Z_j Z_{j+1} - J\sum_{j=1}^{\frac{L_{\rm eff}}{2}} X_j,
    \nonumber \\
    &~H_{\rm int} = \lambda \sum_{j=\frac{L_h}{4}+1}^{\frac{L_h}{4} + \frac{L_b}{2}-2} (X_j Z_{j+1} Z_{j+2} + Z_j Z_{j+1} X_{j+2}),
    \nonumber \\
    &~\delta H = -\delta J \left( X_{\frac{L_h}{4}+1} + X_{\frac{L_h}{4} + \frac{L_b}{2}}\right).
\end{align}

\section{Effective-model level structure from DMRG}
\label{app:levels}

\begin{figure}[t]
	\includegraphics[width=\linewidth]{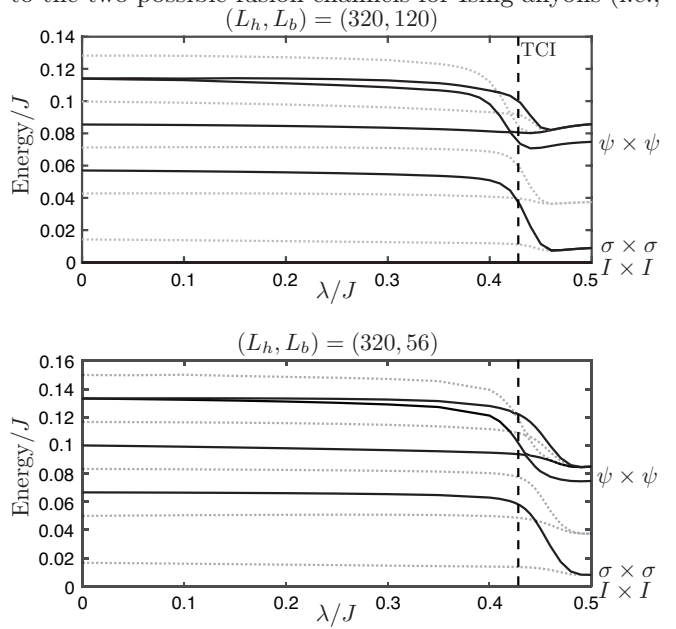}
	\caption{\label{fig:levels} 
	Lowest 10 levels (with the ground state energy subtracted) obtained by DMRG for $(L_h, L_b) = (320, 120)$ and $(320, 56)$. Black curves indicate levels that have trivial total topological charge and thus represent physically accessible states in our spin-liquid problem.  Grey dotted curves corresponds to levels with an odd number of fermions---which we assume are inaccessible in the spin liquid context.
	}
\end{figure}

Figure~\ref{fig:levels} plots the 10 smallest energy eigenvalues of the effective Hamiltonian [Eq.~\eqref{eq:JW}] versus $\lambda$, obtained from DMRG with $(L_h, L_b) = (320, 120)$ and $(320, 56)$; note that the ground state energy has been fixed to zero for each $\lambda$.  
The many-body energies at $\lambda=0$ can be extracted from the theory of a single chiral Majorana fermion propagating over length $L =2L_h + 2L_b$, as arises for the trivial-bridge spin liquid setup that the Hamiltonian emulates in this limit.  That is, the many-body energies follow from summing up single-fermion energies of the form $E_\psi = \frac{2\pi v}{L}(k + 1/2)$ with integer $k \geq 0$.  Dotted grey lines correspond to levels with odd fermion number, while solid black lines have even fermion number.  The former represent states that are not physically accessible in our spin-liquid problem (where we assume trivial total topological charge for the dumbbell).

The levels in Fig.~\ref{fig:levels} at first vary weakly with $\lambda$, but undergo more dramatic evolution near the TCI point ($\lambda = 0.428J$, corresponding to the vertical dashed line).
At $\lambda = 0.5$---where our protocol terminates---the lowest levels that are accessible in the spin liquid problem correspond to the following anyon sectors:  $I\times I$ (ground state), $\sigma\times\sigma$ ($1^{\rm st}$ excitation), $\psi\times\psi$ ($2^{\rm nd}$ excitation), etc.  See the labels on the right side of Fig.~\ref{fig:levels}.  Notice that the first excited state is actually two-fold degenerate, with one of the degenerate levels representing an inaccessible, odd-fermion-parity state.  This pair of levels corresponds to the two possible fusion channels for Ising anyons (i.e., $\sigma \times \sigma = I$ or $\psi$).  In the spin representation, one can identify the states in the $\sigma\times\sigma$ sector by the properties $\braket{Z_j Z_{j+1}}\approx1$ and $\braket{X_j}\approx0$ for sites emulating the bridge region; the $I\times I$ and $\psi\times\psi$ states instead have $\braket{Z_j Z_{j+1}}\approx0$ and $\braket{X_j}\approx1$ in the bridge sites.

Our time-dependent protocol simulations initialize the system into the $\lambda = 0$ ground state. Figure~\ref{fig:levels} indicates that the ground state does not exhibit any level crossings en route to $\lambda = 0.5$; consequently, diabatic evolution is required to access $\sigma\times\sigma$ (and other excited states).  Evolution from the ground state to the $\sigma \times \sigma = \psi$ fusion sector is forbidden even in the spin representation (due to Ising symmetry), and even for diabatic evolution---modulo TEBD simulation errors from truncation and finite time steps.  When computing the probability of entering the $\sigma \times \sigma$ state at the end of the protocol, we use DMRG to compute the pair of degenerate $\sigma \times \sigma$ eigenstates and then sum the probabilities for both. 
The probability of the $\sigma \times \sigma = \psi$ sector due to TEBD error is negligible (e.g., $4\times10^{-8}$ for $L_h=160$, $L_b=80$ and $\tau=1200$).

\end{document}